 \documentclass[final,3p,times]{elsarticle}

\usepackage{amssymb}

\usepackage{lineno}
\usepackage{slashbox}
\usepackage{rotating}

\usepackage{color}
\usepackage{multirow}
\journal{Journal Name}

\newtheorem{thm}{Theorem}

\newdefinition{rmk}{Remark}
\newtheorem{prop}{Proposition}
\newtheorem{coro}{Corollary}
\newproof{pf}{Proof}
\newproof{pot}{Proof of Theorem \ref{thm2}}

\begin{document}

\begin{frontmatter}

\title{On Inference of Overlapping Coefficients in Two Inverse Lomax Populations}

\author[mysecondaryaddress]{Hamza Dhaker\corref{mycorrespondingauthor}}
\cortext[mycorrespondingauthor]{Corresponding author}
\ead{hamza.dhaker@umoncton.ca}
\author[mymainaddress]{El Hadji Deme}
\author[mysecondaryaddress]{and Salah El-Adlouni}
\address[mysecondaryaddress]{D\'{e}partement de math\'{e}matiques et statistique,Universit\'{e} de Moncton, NB, Canada }
\address[mymainaddress]{LERSTAD,UFR SAT, Universite Gaston Berger, Saint-Louis, Senegal}

\begin{abstract}
Overlapping coefficient is a direct measure of similarity between two distributions which is recently becoming very useful.
This paper investigates estimation for some well-known measures of overlap, namely Matusita's measure $\rho$,  Weitzman's measure $\Delta$ and $\Lambda$ based on Kullback-Leibler. 
Two estimation methods considered in this study are point estimation and Bayesian approach. Two Inverse Lomax populations with different shape parameters are considered. The bias and mean square error properties of the estimators are studied through a simulation study  and a real data example.
\end{abstract}

\end{frontmatter}


\section{Introduction}
Inverse Lomax distribution is a special case of the Generalized Beta distribution of the second kind. It is one of the notable lifetime models in statistical applications. The inverse Lomax distribution is one of significant lifetime models. Kleiber \cite{Kleiber} used this Inverse Lomax distribution to get Lorenz ordering relationship among ordered statistics. McKenzie et al. \cite{McKenzie} applied this life time distribution  on geophysical data on the sizes of land fires in the California state, US.\\

 The Overlapping Coefficients ($OVL$) represents the proportion of overlap between two probability density functions (pdf) as a measure of similarity between distributions.
Generally it is measured on the scale of 0 to 1. Values of measure close to 0 corresponding to the distributions having supports with no intersection and 1 to the perfect matching of the two distributions. 
This paper investigates point and interval estimation for four measures of overlap (OVL) for two Inverse Lomax populations with different shape Parameters. 
\begin{itemize}
\item[$\bullet$] Matusia's Measure \cite{Matusita} $$ \rho = \int\sqrt{f_{1}(x)f_{2}(x)}dx $$
\item[$\bullet$] Weitzman's Measure \cite{Weitzman} $$ \Delta = \int \min \lbrace f_{1}(x),f_{2}(x)\rbrace dx $$
\item[$\bullet$] OVL based Kullback-Leibler \cite{Kullback}
\begin{eqnarray}
\label{kll}
 \Lambda = \frac{1}{1+KL(f_{1}\| f_{2} )}
 \end{eqnarray}
with $KL(f_{1}\| f_{2})= \int (f_{1}(x)-f_{2}(x))\log \left(\frac{f_{1}(x)}{f_{2}(x)} \right)dx  $
\end{itemize}

The mathematical structure of these measures is complicated; there are no results available on the exact sampling distributions of the commonly used OVL estimators. Researchers such as Smith \cite{Smith} derived formulas for estimating the mean and the variance of discrete version of Weitzman’s measure using the delta method. Mishra et al. \cite{Mishra} gave small and large sample properties of the sampling distribution for a function of $\Delta$ under the assumption of homogeneity of variances.
Recently, several authors including Al-Saidy et al. \cite{Al-Saidy},  Bradley \cite{Bradley}, Clemons \cite{Clemons}, Dhaker et al. \cite{Dhaker}, Inman \cite{Inman}, Jose \cite{Jose}, Mulekar  \cite{Mulekar} and Reiser \cite{Reiser}  considered this measure.
 \\

In this article, we consider the point and interval estimation for some measures of overlap (OVL) for two Inverse Lomax populations with different shape Parameters using "Simple Random Sample (SRS) and Ranked Set Sampling (RSS) and Bayesian methodology". \\ 
The first method (RSS , McIntyre \cite{McIntyre}) was earlier applied by Helu and Samawi \cite{Helu} for the point and interval estimation of the overlapping coefficients for two Lomax distributions. We will use their methodology for the point estimate and interval in the case of inverse Lomax distribution. The second approach, we use another method for parameter estimation using Bayesian inference \cite{Jan}.
\\
The primary purpose of this study is to compare the confidence intervals for the overlap  coefficients ($\rho$, $\Delta$ and $\Lambda$) computed using SRS, RSS, Bayesian methods. 
Section 2 defines the inverse Lomax distribution and derivations of the three OVL measures. In Section 3 we draw some statistical inference on the OVL measures using SRS. Section 4 draws an inference on the 
OVL measures using RSS.  In Section 5, we provide Bayesian estimators along with approximate bias and variances for the three measures of overlap. In Section 6, a simulation study is performed to evaluate and compare biases and mean square errors of OVL measures estimates. In Section 7 we give an example using a real dataset. Finally, the conclusion is presented in Section 8.

\begin{figure}
\begin{center}
\label{exponentiel}
\includegraphics[scale=0.5]{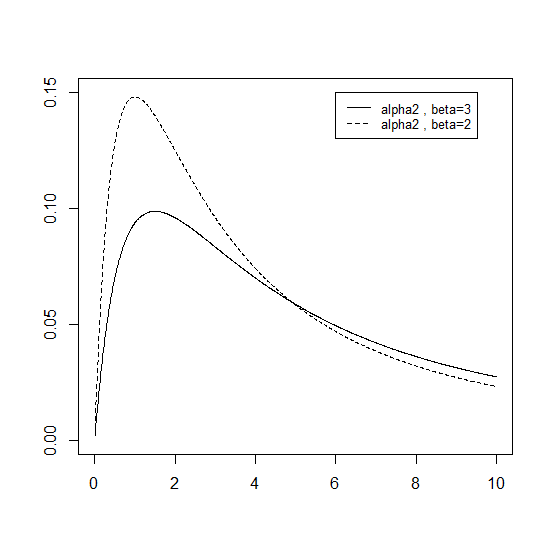}
\end{center}
\caption{The overlap of two inverse Lomax densities.}
\end{figure}

\section{OVL measures for inverse Lomax distribution}
 A random variable $X$ is said to have a Lomax distribution if the corresponding probability density function and cumulative density function are given by Yadav et al \cite{Yadav}.

 \begin{eqnarray}
  g(y;\alpha ,\beta)=\frac{\alpha}{\beta}\left(1+\frac{y}{\beta}\right)^{-(\alpha +1)} \qquad  y\geq0, \quad \alpha \quad  ,\beta > 0 
 \end{eqnarray}

 \begin{eqnarray}
  G(y;\alpha ,\beta)=1-\left(1+\frac{y}{\beta}\right)^{-\alpha} \qquad  y\geq0, \quad \alpha \quad  ,\beta > 0
 \end{eqnarray}

Consider the random variable $Z=\frac{1}{Y}$. Then $Z$ has the inverse Lomax distribution with $pdf$ and $cdf$ as

\begin{eqnarray}
h(z;\alpha ,\beta)=\frac{\beta}{\alpha z^{2}} \left(1+\frac{\beta}{z}\right)^{-(1+1/\alpha)} \qquad  z\geq0, \alpha ,\beta > 0 
 \end{eqnarray}
\begin{eqnarray}
H(z;\alpha ,\beta)=\left(1+\frac{\beta}{z} \right)^{-1/\alpha} \qquad  z\geq0, \alpha ,\beta > 0
\end{eqnarray}

respectively. Note that $h(y;\alpha ,\beta)=\frac{1}{z^{2}}f(\frac{1}{z})$ and $H(y;\alpha ,\beta)=1-F(\frac{1}{z})$.
\\

We consider another variable with $X=\frac{Z}{\beta}$

\begin{eqnarray}
\label{inl}
f(x;\alpha)=\frac{1}{\alpha x^{2}} \left(1+\frac{1}{x}\right)^{-(1+1/\alpha)} \qquad  x\geq0, \alpha > 0 
 \end{eqnarray}
\begin{eqnarray}
F(x;\alpha)=\left(1+\frac{1}{x} \right)^{1-1/\alpha} \qquad  x\geq0, \alpha  > 0
\end{eqnarray}

The computation or estimation of OVL for two inverse Lomax distributions, with density functions :

\begin{eqnarray}
f_1(x;\alpha_1)=\frac{1}{\alpha_1 x^{2}} \left(1+\frac{1}{x}\right)^{-(1+1/\alpha_1)} \qquad  x\geq0, \alpha_1 > 0 
 \end{eqnarray}
 \begin{eqnarray}
f_2(x;\alpha_2)=\frac{1}{\alpha_2 x^{2}} \left(1+\frac{1}{x}\right)^{-(1+1/\alpha_2)} \qquad  x\geq0, \alpha_2 > 0 
 \end{eqnarray}
 
 Let $R=\frac{\alpha_{1}}{\alpha_{2}}$, the continuous version of the three overlap measures can be expressed as a function of $C$ as follows:

 \begin{eqnarray}
\rho = \frac{2\sqrt{R}}{R+1},
 \end{eqnarray}

 \begin{eqnarray}
\label{delta}
\Delta = 1-R^{\frac{1}{1-R}}\mid 1-\frac{1}{R}\vert , \qquad R\neq 1,
\end{eqnarray}
and
 \begin{eqnarray}
\Lambda= \frac{R}{R^{2}-R+1}.
 \end{eqnarray}

\begin{prop}
For OVLs defined earlier,
\begin{itemize}
\item[i)] $0\leq OVL\leq 1$ for all $R\geq 0$
\item[ii)] $ OVL = 1$ iff $R=1$
\item[iii)]  $ OVL = 0$ iff $R=0$ or $R=\infty$.
\end{itemize}
\end{prop}

\begin{prop}
All four OVLs possess properties of reciprocity, invariance, and piecewise monotonicity
\begin{itemize}
\item[i)] $OVL(R)= OVL(1/R)$,
\item[ii)] $OVLs$ are monotonically increasing in $R$ for $0\leq R\leq 1$ and decreasing in
 $R> 1.$
\end{itemize}
\end{prop}

\begin{figure}
\begin{center}
\label{exponentiel}
\includegraphics[scale=0.5]{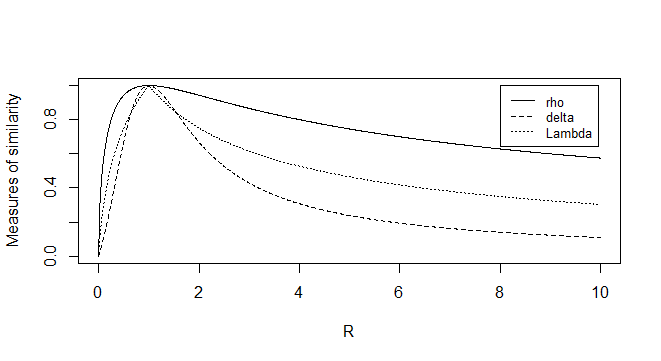}
\end{center}
\caption{Measures of similarity as functions of R.}
\end{figure}

\section{Statistical inference using Simple Random Sample}
\subsection{Estimation}
\label{seces}
As in Helu and Samawi \cite{Helu}, parallel results to those of the two Lomax populations can be established for the  inverse Lomax populations.\\
Suppose $(X_{ij}; j= 1,...,n_i; i = 1,2)$ denote independent observations from two independent inverse Lomex populations. 
Let $f_i(x) (i=1,2)$ denote the inverse Lomex densities with shape parameters $\alpha_1$ and $\alpha_2$ respectively. Define $R=\alpha_1/\alpha_2$.
The likelihood function the Inverse Lomax distribution (\ref{inl}) is given as:
\begin{eqnarray}
\label{lik}
L(\alpha_i\vert x)=\frac{1}{\alpha_i^{n}} \prod_{j=1}^{n}\frac{1}{x_{j}^{2}}\prod_{j=1}^{n}ln\left( 1+\frac{1}{x_j} \right) \qquad i=1,2;j=1,...n.
\end{eqnarray}
 The maximum likelihood estimators (MLEs) based on the two samples are given by:
\begin{itemize}
\item[1-]  From the first sample:
$$ \widehat{\alpha}_{1SRS}=\frac{1}{n_1}\sum_{j=1}^{n_{1}}\log(1+\frac{1}{x_{1j}}) $$
\item[2-]  From the second sample:
$$ \widehat{\alpha}_{2SRS}=\frac{1}{n_2}\sum_{j=1}^{n_{2}}\log(1+\frac{1}{x_{2j}}) $$
\end{itemize}

The maximum likelihood estimators $\widehat{\alpha}_{SRS1}$ and $\widehat{\alpha}_{SRS2}$ exist and are unique. Using a simple transformation, it can be shown that 
$$ \widehat{\alpha}_{SRS1} \sim Gamma(n_1,\frac{\alpha_1}{n_1}) \quad and \quad   \widehat{\alpha}_{SRS2} \sim Gamma(n_2,\frac{\alpha_2}{n_2})  $$
Consequently, the means and variances of those $MLE's$ are respectively
$$ \mathbb{E}(\widehat{\alpha}_{SRS1}) =\alpha_{1} \qquad \mathbb{E}(\widehat{\alpha}_{2}) =\alpha_{2},$$
and
$$ V(\widehat{\alpha}_{SRS1})=\frac{\alpha_{1}^2}{n_1} \qquad  V(\widehat{\alpha}_{SRS2})=\frac{\alpha_{1}^2}{n_2} .$$

Then we may define an estimate of $R$ is
$$ \widehat{R}=\frac{\widehat{\alpha}_{\widehat{\alpha}1}}{\widehat{\alpha}_{\widehat{\alpha}2}}. $$

Therefore, using the relationship 
between Gamma distribution and Chi-
square distribution and the fact that 
the two samples are independent, it is 
easy to show that $\frac{\alpha_{2}}
{\alpha_{1}}\widehat{R}$ has $F$-
distribution with $2n_{1}$ and  $2n_{2}$ 
degrees of freedom ($F_{2n_1,2n_2}$). 
Hence, the variance of $\widehat{R}$ is 
$$Var(\widehat{R}_{SRS})=\frac{n_{2}^{2}
(n_1+n_2 -1)}{n_1(n_2-1)^{2}(n_{2}-2)}
R^2 .$$
Also, an unibiased estimate $R$ is given by $\widehat{R}_{SRS}^{*}=\frac{n_{2}-1}{n_{2}}\widehat{R}_{SRS}$ with

$$Var(\widehat{R}^{*}_{SRS})=R^2 \frac{n_1+n_2-1}{n_{1}(n_2-2)},$$
Clearly, $\widehat{R}^{*}_{SRS}$ has less variance than $\widehat{R}_{SRS}$.\\
Since the $OVL$ measures are functions of $R$, therefore, based on $MLE$ estimate of $R$, the $OVL$ measures can be estimated by

 \begin{eqnarray}
\widehat{\rho}_{SRS} = \frac{2\sqrt{\widehat{R}^{*}_{SRS}}}{\widehat{R}^{*}_{SRS}+1},
 \end{eqnarray}

 \begin{eqnarray}
\label{delta}
\label{delta}
\Delta = 1-(\widehat{R}^{*})^{\frac{1}{1-\widehat{R}^{*}}}\mid 1-\frac{1}{\widehat{R}^{*}}\vert , \qquad R\neq 1,
\end{eqnarray}
and
 \begin{eqnarray}
\widehat{\Lambda}_{SRS}= \frac{\widehat{R}^{*}_{SRS}}{(\widehat{R}^{*}_{SRS})^{2}-\widehat{R}^{*}_{SRS}+1}.
 \end{eqnarray}

\subsection{Asymptotic properties}
Let $OVL= g(R)$, and its estimator $\widehat{OVL}_{SRS}= g(\widehat{R}^{*}_{SRS})$.
Using the well-known Delta method (expansion of the Taylor series) the asymptotic sampling variance of the $ OVL $ measures is given by the following theorem

\begin{thm}
\label{thm1}
Let $\widehat{\rho}_{SRS}$, $\widehat{\Delta}_{SRS}$ and $\widehat{\Lambda}_{SRS}$ are the estimates of $\rho$, $\Delta$ and $\Lambda$ respectively, then for $n_2\geq 3$, we have the approximate expressions for variances of the $OVL$ measures can be obtained as follows:

$$ Var(\widehat{\rho}_{SRS})=\frac{(n_1+n_2-1)}{n_1(n_2-2)}\frac{R(1-R)^{2}}{(R+1)^{4}}, $$

$$ Var(\widehat{\Delta}_{SRS}) = \frac{(n_1+n_2-1)}{n_1(n_2-2)}\frac{R^{\frac{2}{1-R}}(lnR)^{2}}{(1-R)^{2}},
 $$
 $$Var(\widehat{\Lambda}_{SRS})=\frac{(n_1+n_2-1)}{n_1(n_2-2)}\frac{R^{2}(1-R^2)^{2}}{(R^{2}-R+1)^{4}}. $$

\end{thm}
\begin{pf}
Let function $g(R)$ has one parameter  of $R$ and let $\widehat{R}^{*}_{SRS}$ be an almost sure consistent estimate of $R$.

 Then the variance of $g(\widehat{R}^{*}_{SRS})$ may be obtained from the linear Taylor approximation around $R$.
$$ g(\widehat{R}^{*}_{SRS})=g(R)+(\widehat{R}^{*}_{SRS}-R)g'(R)$$ 

for the estimator $\widehat{\rho}_{SRS}$:
$$g(\widehat{R}^{*}_{SRS})= \frac{2\sqrt{\widehat{R}^{*}_{SRS}}}{\widehat{R}^{*}_{SRS}+1}$$
Since, in this case, $$ g'(\widehat{R})= \frac{1-\widehat{R}}{\sqrt{\widehat{R}}(1+\widehat{R})^{2}} $$

 \begin{eqnarray*}
Var(\widehat{\rho}_{SRS})&=& Var(g(\widehat{R}^{*}_{SRS}))=Var(g(R))+Var((\widehat{R}^{*}_{SRS}-R)g'(R)) \\ &=& (g'(R))^{2}Var(\widehat{R}^{*}_{SRS}) = \frac{(1-R)^2}{R(1+R)^4}R^2\frac{n_1+n_2-1}{n_1(n_2-1)} \\&=& \frac{(n_1+n_2-1) }{n_1(n_2-2)}\frac{R(1-R)^2}{(1+R)^4}
\end{eqnarray*} 

Similar arguments can be used for the other overlaps coefficients.
\end{pf}

\begin{thm}
\label{thm2}
Using Taylor series expansion, then for $n_2\geq 3$. Approximations for the biases of the OVL coefficients estimates are as follows:
$$ Bias(\widehat{\rho}_{SRS})= \frac{(n_1+n_2-1)}{2n_1(n_2-2)}\frac{\sqrt{R}(3R^2-6R-1)}{(1+R)^{3}}  $$

$$ Bias(\widehat{\Delta}_{SRS}) = \displaystyle\left\{
    \begin{array}{l}
  -\frac{(n_1+n_2-1)}{2n_1(n_2-2)}R^{2}\left[  \frac{  R^{\frac{2R-1}{1-R} }R( 2R-lnR-2)lnR-(R-1)^{2} }{(R-1)^{3}} \right]
     \qquad if \quad 0<R<1 \\
     \\
 \frac{(n_1+n_2-1)}{2n_1(n_2-2)}R^{2}\left[  \frac{  R^{\frac{2R-1}{1-R} }R( 2R-lnR-2)lnR-(R-1)^{2} }{(R-1)^{3}} \right]   
     \qquad if \quad R\geq 1
 \end{array}
\right.
 $$

$$Bias(\widehat{\Lambda}_{SRS})= \frac{(n_1+n_2-1)}{n_1(n_2-2)}\frac{R^5-3R^3-R^2}{(R^2-R+1)^2} $$

\end{thm}
\begin{pf}
Again by using the well-known Delta method (Taylor series expansion) the asymptotic bias of the OVL measures can be obtained as follows:
$$ g(\widehat{R}^{*}_{SRS})=g(R)+(\widehat{R}^{*}_{SRS}-R)g'(R)+\frac{1}{2}(\widehat{R}^{*}_{SRS}-R)^2g''(R)$$ 
for the estimator $\widehat{\rho}_{SRS}$:
$$g(\widehat{R}^{*}_{SRS})= \frac{2\sqrt{\widehat{R}^{*}_{SRS}}}{\widehat{R}^{*}_{SRS}+1}$$
Since, in this case, $$ g'(\widehat{R}^{*}_{SRS})= \frac{1-\widehat{R}^{*}_{SRS}}{\sqrt{\widehat{R}^{*}_{SRS}}(1+\widehat{R}^{*}_{SRS})^{2}} \quad so \quad g''(\widehat{R}^{*}_{SRS})= \frac{\sqrt{R}(3R^{2}-6R-1)}{R^{3/2}(1+R)^3} $$

 \begin{eqnarray*}
\mathbb{E}(\widehat{\rho}_{SRS})&=& \mathbb{E}(g(\widehat{R}^{*}_{SRS}))=g(R)+\mathbb{E}\left[(\widehat{R}^{*}_{SRS}-R)\right]g'(R)+\frac{1}{2}\mathbb{E}\left[(\widehat{R}^{*}_{SRS}-R)^2\right] g''(R) \\ Bias(\widehat{\rho}_{SRS}) &=& \frac{1}{2}Var(\widehat{R}^{*}_{SRS})g''(R) \\ &=& \frac{(n_1+n_2-1)}{2n_1(n_2-2)}\frac{\sqrt{R}(3R^2-6R-1)}{(1+R)^{3}} 
\end{eqnarray*} the bias
 Similar arguments can be used for the bias the other overlaps coefficients.
\end{pf}

\subsection{Interval estimation}
For large sample, normal approximation to the sampling distribution, using the Delta-method, works fairly well. Therefore, the asymptotic $100(1-\alpha_0)\%$ confidence intervals for the OVL coefficients can be computed easily as:
$$ \left\lbrace \widehat{OVL}_{SRS} \quad \underline{+} \quad Z_{1-\alpha_0/2}\sqrt{Var(\widehat{OVL}_{SRS})} \right\rbrace $$
where $Z_{1-\alpha_0/2}$ is the $\alpha_0/2$ upper quantile of the standard normal distribution.
\\

These confidence intervals are not the best because of the bias involved in $OVL$ coefficients estimates, however, for large samples they work fairly well. Using these approximations, the bias corrected interval can be computed as

$$ \left\lbrace \left[ \widehat{OVL}_{SRS}-Bias(\widehat{OVL}_{SRS})\right] \quad \underline{+} \quad Z_{1-\alpha_0/2}\sqrt{Var(\widehat{OVL}_{SRS})} \right\rbrace $$

\section{Statistical inference using Ranked Set Sampling}
\subsection{Estimation}
Similar to the previous section, suppose $(X_{1(1)k},X_{1(2)k},...,X_{1(r_1)k})$ 
and $(X_{2(1)k},X_{2(2)k},...,X_{2(r_1)k})$, $k=1,2,...,m$ are two independent $RSS$ samples drawn from $f_{1}(x)$ and $f_2(x)$ respectively. The estimates of $\theta_{1}$ and $\theta_{2}$ using $RSS$ sample are given by:

\begin{itemize}
\item[1-]  From the first sample:
$$ \widehat{\alpha}_{1RSS}=\frac{1}{n_1}\sum_{i=1}^{r_1}\sum_{k=1}^{n_{1}}\log(1+\frac{1}{x_{1(i)k}}), \qquad n_1=r_1m. $$
\item[2-]  From the second sample:
$$ \widehat{\alpha}_{2RSS}=\frac{1}{n_2}\sum_{i=1}^{r_2}\sum_{k=1}^{n_{2}}\log(1+\frac{1}{x_{2(i)k}}), \qquad n_2=r_2m. $$
\end{itemize}
Note that, it is easy to show that

$$ \mathbb{E}(\widehat{\alpha}_{1RSS})=\alpha_{1}, \quad \mathbb{E}(\widehat{\alpha}_{2RSS})=\alpha_{2},$$ 
$$Var(\widehat{\alpha}_{1RSS})=\frac{\alpha_{1RSS}^{2}}{mr_{1}^{2}}\sum_{i=1}^{r_{1}}\frac{1}{r_1-i+1} \quad  Var(\widehat{\alpha}_{2RSS})=\frac{\alpha_{2RSS}^{2}}{mr_{2}^{2}}\sum_{i=1}^{r_{2}}\frac{1}{r_2-i+1}$$
Also, $R$ can be estimated by $\widehat{R}_{RSS}=\frac{\widehat{\alpha}_{1RSS}}{\widehat{\alpha}_{2RSS}}$.
Hence, by using Delta method of approximation, the variance of $\widehat{R}_{RSS}$ can be approximated by
$$ Var(\widehat{R}_{RSS})\cong R^{2}
\left[  \frac{  \sum_{i=1}^{r_{1}}\frac{1}{r_1-i+1} }{mr_{1}^{2}} + \frac{  \sum_{i=1}^{r_{2}}\frac{1}{r_2-i+1} }{mr_{2}^{2}} \right]. $$
Thus, we have
 \begin{eqnarray}
\widehat{\rho}_{RSS} = \frac{2\sqrt{\widehat{R}^{*}_{RSS}}}{\widehat{R}^{*}_{RSS}+1}
 \end{eqnarray}

 \begin{eqnarray}
\label{delta}
\widehat{\Delta}_{RSS} = \displaystyle\left\{
    \begin{array}{l}
    1- (\widehat{R}^{*}_{RSS})^{\frac{1}{\widehat{R}^{*}_{RSS}-1}}+(\widehat{R}^{*}_{RSS})^{\frac{\widehat{R}^{*}_{RSS}}{\widehat{R}^{*}_{RSS}-1}} \qquad if \quad 0<R<1 \\
     \\
    1+ (\widehat{R}^{*}_{RSS})^{\frac{1}{\widehat{R}^{*}_{RSS}-1}}-(\widehat{R}^{*}_{RSS})^{\frac{\widehat{R}^{*}_{RSS}}{\widehat{R}^{*}_{RSS}-1}} \qquad if \quad R\geq 1
 \end{array}
\right.
\end{eqnarray}

 \begin{eqnarray}
\widehat{\Lambda}_{RSS}= \frac{\widehat{R}^{*}_{RSS}}{(\widehat{R}^{*}_{RSS})^{2}-\widehat{R}^{*}_{RSS}+1}
 \end{eqnarray}

\subsection{Asymptotic properties}
Let $OVL= g(R)$, and its estimator $\widehat{OVL}_{RSS}= g(\widehat{R}_{RSS})$.
Using the well-known Delta method (expansion of the Taylor series) the asymptotic sampling variance of the $ OVL $ measures is given by the following theorem

\begin{coro}
\label{thm3}
Let $\widehat{\rho}_{RSS}$, $\widehat{\Delta}_{RSS}$ and $\widehat{\Lambda}_{RSS}$ are the estimates of $\rho$, $\Delta$ and $\Lambda$ respectively, then for $n_2\geq 3$, we have the approximate expressions for variances of the $OVL$ measures can be obtained as follows:

$$ Var(\widehat{\rho}_{RSS})= \left[  \frac{  \sum_{i=1}^{r_{1}}\frac{1}{r_1-i+1} }{mr_{1}^{2}} + \frac{  \sum_{i=1}^{r_{2}}\frac{1}{r_2-i+1} }{mr_{2}^{2}} \right]\frac{R(1-R)^{2}}{(R+1)^{4}} $$

$$ Var(\widehat{\Delta}_{RSS}) =  \left[  \frac{  \sum_{i=1}^{r_{1}}\frac{1}{r_1-i+1} }{mr_{1}^{2}} + \frac{  \sum_{i=1}^{r_{2}}\frac{1}{r_2-i+1} }{mr_{2}^{2}} \right]\frac{R^{\frac{2}{1-R}}(lnR)^{2}}{(1-R)^{2}}
 $$
 $$Var(\widehat{\Lambda}_{RSS})=\left[  \frac{  \sum_{i=1}^{r_{1}}\frac{1}{r_1-i+1} }{mr_{1}^{2}} + \frac{  \sum_{i=1}^{r_{2}}\frac{1}{r_2-i+1} }{mr_{2}^{2}} \right]\frac{R^{2}(1-R^2)^{2}}{(R^{2}-R+1)^{4}} $$

\end{coro}
\begin{pf}
same proof of Theorem\ref{thm1}, replacing $\widehat{R}^{*}_{RRS}$ with the $\widehat{R}^{*}_{SRS}$ estimator.
\end{pf}

\begin{coro}
\label{thm4}
Using Taylor series expansion, then for $n_2\geq 3$. Approximations for the biases of the OVL coefficients estimates, are as follow:
$$ Bias(\widehat{\rho}_{RSS})= \left[  \frac{  \sum_{i=1}^{r_{1}}\frac{1}{r_1-i+1} }{mr_{1}^{2}} + \frac{  \sum_{i=1}^{r_{2}}\frac{1}{r_2-i+1} }{mr_{2}^{2}} \right]\frac{\sqrt{R}(3R^2-6R-1)}{2(1+R)^{3}}  $$

$$ Bias(\widehat{\Delta}_{RSS}) = \displaystyle\left\{
    \begin{array}{l}
  -\left[  \frac{  \sum_{i=1}^{r_{1}}\frac{1}{r_1-i+1} }{mr_{1}^{2}} + \frac{  \sum_{i=1}^{r_{2}}\frac{1}{r_2-i+1} }{mr_{2}^{2}} \right]R^{2}\left[  \frac{  R^{\frac{2R-1}{1-R} }R( 2R-lnR-2)lnR-(R-1)^{2} }{(R-1)^{3}} \right]
     \qquad if \quad 0<R<1 \\
     \\
\left[  \frac{  \sum_{i=1}^{r_{1}}\frac{1}{r_1-i+1} }{mr_{1}^{2}} + \frac{  \sum_{i=1}^{r_{2}}\frac{1}{r_2-i+1} }{mr_{2}^{2}} \right]R^{2}\left[  \frac{  R^{\frac{2R-1}{1-R} }R( 2R-lnR-2)lnR-(R-1)^{2} }{(R-1)^{3}} \right]  
     \qquad if \quad R\geq 1
 \end{array}
\right.
 $$

$$Bias(\widehat{\Lambda}_{RSS})= \left[  \frac{  \sum_{i=1}^{r_{1}}\frac{1}{r_1-i+1} }{mr_{1}^{2}} + \frac{  \sum_{i=1}^{r_{2}}\frac{1}{r_2-i+1} }{mr_{2}^{2}} \right]\frac{R^5-3R^3-R^2}{(R^2-R+1)^2} $$

\end{coro}
\begin{pf}
same proof of Theorem\ref{thm2}, replacing $\widehat{R}^{*}_{RRS}$ with the $\widehat{R}^{*}_{SRS}$ estimator.
\end{pf}

\subsection{Interval estimation}
Similar to the case of SRS and RSS, the asymptotic $100(1-\alpha_0)\%$ confidence intervals for the OVL coefficients can be computed easily as:
$$ \left\lbrace \widehat{OVL}_{RSS} \quad \underline{+} \quad Z_{1-\alpha_0/2}\sqrt{Var(\widehat{OVL}_{RSS})} \right\rbrace $$
where $Z_{1-\alpha_0/2}$ is the $\alpha_0/2$ upper quantile of the standard normal distribution.
\\

These confidence intervals are not the best because of the bias involved in $OVL$ coefficient estimates, however, for large samples they work fairly well. Using these approximations, the bias corrected interval can be computed as

$$ \left\lbrace \left[ \widehat{OVL}_{RSS}-Bias(\widehat{OVL}_{RSS})\right] \quad \underline{+} \quad Z_{1-\alpha/2}\sqrt{Var(\widehat{OVL}_{RSS})} \right\rbrace $$


\section{Statistical inference using Bayesian Approach}
In recent decades, the Bayes viewpoint, as a powerful and valid alternative to traditional statistical perspectives, has received frequent attention for statistical inference.
In our study normal approximations for the shape parameter $\alpha$ of Inverse Lomax distribution will be obtained using Jeffery's prior. Noted that the choice of this type of distribution, thus often leads to classical estimators of the maximum likelihood approach. 
\subsection{Estimation}
\begin{itemize}
\item[-]\textbf{Jeffery's Prior:}
Using Jeffery's prior for the scale parameter $\alpha$
\begin{eqnarray}
\label{priori1}
P(\alpha)= \alpha^{-1} \qquad  0<\alpha <\infty
\end{eqnarray}

Using (\ref{priori1}) and (\ref{lik}) we get the posterior distribution 
for $\alpha$ is as:

\begin{eqnarray}
P(\alpha\vert  x) \propto P(\alpha) L(\alpha\vert x)=\frac{1}{\alpha^{n+1}}\exp\left(-(1+1/\alpha)\sum_{i=1}^{n}\log(1+\frac{1}{x_{i}}) \right)
\end{eqnarray}
The log posterior is $\log(P(\alpha\vert  x))=-(n+1)\log(\alpha)-(1+1/\alpha)\sum_{i=1}^{n}\log(1+\frac{1}{x_{i}})$ \\
The first derivative is
$$\frac{\partial P(\alpha\vert  x)}{\partial \alpha}= -\frac{n+1}{\alpha}+\frac{1}{\alpha^{2}} \sum_{i=1}^{n}\log(1+\frac{1}{x_{i}})$$
and the posterior mode is obtained as:

\begin{itemize}
\item[1-]  From the first sample:
$$ \widehat{\alpha}_{1J}=\frac{1}{n_1+1}\sum_{i=1}^{n_1}\log(1+\frac{1}{x_{1i}}) $$
\item[2-]  From the second sample:
$$ \widehat{\alpha}_{2J}=\frac{1}{n_2+1}\sum_{i=1}^{n_2}\log(1+\frac{1}{x_{2i}}) $$
\end{itemize}
Using simple transformation, it can be shown that
$$ \widehat{\alpha}_{1J} \sim Gamma(n_1,\frac{\alpha_1}{n_1 +1}) \quad and \quad   \widehat{\alpha}_{2J} \sim Gamma(n_2,\frac{\alpha_2}{n_2+1})  $$
 A consequent estimate of
$R$ is $\widehat{R}_{J}=\frac{ \widehat{\alpha}_{1J}}{ \widehat{\alpha}_{2J}}$. Hence, an approximation variance of $\widehat{R}_{J}$ can be given by
$$ Var(\widehat{R}_{J})=\left( \frac{n_2+1}{n_1+1}\right)^{2}\frac{n_{2}^{4}(n_1+n_2-1)}{n_{1}^{3}(n_2-1)^{2}(n_2-2)}R^2 $$

Also, an unibiased estimate $R$ is given by $\widehat{R}_{J}^{*}=\frac{n_1(n_1-1)(n_1+1)}{n_{2}^{2}(n_{2}+1)}\widehat{R}_{BJ}$ with
$$ Var(\widehat{R}_{J}^{*})=\left(\frac{n_1-1}{n_2-1}\right)^2\frac{n_1+n_2-1}{n_1(n_2-2)}R^2 $$
Thus, we have

 \begin{eqnarray}
\widehat{\rho}_{J} = \frac{2\sqrt{\widehat{R}^{*}_{J}}}{\widehat{R}^{*}_{J}+1}
 \end{eqnarray}
 \begin{eqnarray}
\label{delta}
\widehat{\Delta}_{J} = \displaystyle\left\{
    \begin{array}{l}
    1- (\widehat{R}^{*}_{J})^{\frac{1}{\widehat{R}^{*}_{J}-1}}+(\widehat{R}^{*}_{J})^{\frac{\widehat{R}^{*}_{J}}{\widehat{R}^{*}_{J}-1}} \qquad if \quad 0<R<1 \\
     \\
    1+ (\widehat{R}^{*}_{J})^{\frac{1}{\widehat{R}^{*}_{J}-1}}-(\widehat{R}^{*}_{J})^{\frac{\widehat{R}^{*}_{J}}{\widehat{R}^{*}_{J}-1}} \qquad if \quad R\geq 1
 \end{array}
\right.
\end{eqnarray}

 \begin{eqnarray}
\widehat{\Lambda}_{J}= \frac{\widehat{R}^{*}_{J}}{(\widehat{R}^{*}_{J})^{2}-\widehat{R}^{*}_{J}+1}
 \end{eqnarray}

The asymptotic variance of the $OVL$ measures are given by:

$$
Var(\widehat{\rho}_{J})= \left( \frac{n_1-1}{n_2-1}\right)^{2}\frac{ n_1+n_2-1}{n_1(n_2-2)}\frac{R(1-R)^2}{(1+R)^4}
$$

$$ Var(\widehat{\Delta}_{J}) = \left( \frac{n_1-1}{n_2-1}\right)^{2}\frac{ n_1+n_2-1}{n_1(n_2-2)} \frac{R^{\frac{2}{1-R}}(lnR)^{2}}{(1-R)^{2}}$$

$$Var(\widehat{\Lambda}_{J})=\left( \frac{n_1-1}{n_2-1}\right)^{2}\frac{ n_1+n_2-1}{n_1(n_2-2)} \frac{R^2(1-R^2)^2}{(R^2-R+1)^4}. $$

With the asymptotic bias given by: 

$$ Bias(\widehat{\rho}_{J})= \left( \frac{n_2+2}{n_1+1}\right)^{2}\frac{n_1(n_1+n_2-1)}{2(n_2-1)^{2}(n_2-2)}\frac{\sqrt{R}(3R^2-6R-1)}{(1+R)^{3}} $$

$$ Bias(\widehat{\Delta}_{J}) = \displaystyle\left\{
    \begin{array}{l}
-\left( \frac{n_2+2}{n_1+1}\right)^{2}\frac{n_1(n_1+n_2-1)}{(n_2-1)^{2}(n_2-2)}R^{2}\left[  \frac{  R^{\frac{2R-1}{1-R} }R( 2R-lnR-2)lnR-(R-1)^{2} }{(R-1)^{3}} \right]
     \qquad if \quad 0<R<1 \\
     \\
\left( \frac{n_2+2}{n_1+1}\right)^{2}\frac{n_1(n_1+n_2-1)}{(n_2-1)^{2}(n_2-2)}R^{2}\left[  \frac{  R^{\frac{2R-1}{1-R} }R( 2R-lnR-2)lnR-(R-1)^{2} }{(R-1)^{3}} \right]   
     \qquad if \quad R\geq 1
 \end{array}
\right.
 $$

$$Bias(\widehat{\Lambda}_{J})= \left( \frac{n_2+2}{n_1+1}\right)^{2}\frac{n_1(n_1+n_2-1)}{(n_2-1)^{2}(n_2-2)}\frac{R^5-3R^3-R^2}{(R^2-R+1)^2} $$

\end{itemize}

\subsection{Interval estimation}
 The $(1-2\alpha_0)$ confidence intervals for the overlap measures are computed as:
$$ \left\lbrace \widehat{OVL}_{J} \quad \underline{+} \quad Z_{1-\alpha_0/2}\sqrt{Var(\widehat{OVL}_{J})} \right\rbrace $$.

 Using these estimates the bias corrected interval, the $100(1-\alpha_0)\%$ confidence intervals for the $OVL$ measures can be given by
$$ \left\lbrace \left[ \widehat{OVL}_{J}-Bias(\widehat{OVL}_{J})\right] \quad \underline{+} \quad Z_{1-\alpha_0/2}\sqrt{Var(\widehat{OVL}_{J})} \right\rbrace $$.

\section{Simulation}
In our simulation study we include the following: $R=0.1, 0.5, 0.75, 0.8$, and $r_1=2,3,4,5$; $r_2=2,3,4,5$; $m =8,40$ and $\alpha_0 =0.05$. A simulation study is conducted to get insight about the performance of the proposed estimators.
All the 1000 simulated sets of observations were generated under the assumption that both densities have standard inverse Lomax distribution with the different sharpe parameter.
\\

The performance of the $OVL$ measure using $RSS$ and $SRS$ can be assessed using the asymptotic relative efficiency which is computed as
$$ Eff(\widehat{OVL}_{SRS},\widehat{OVL}_{RSS})=\frac{MSE(\widehat{OVL}_{SRS})}{MSE(\widehat{OVL}_{RSS})} $$
Where $MSE(\widehat{OVL})=Var(\widehat{OVL})+Bias(\widehat{OVL})^2$
\\
Tables 1 and 2 show the asymptotic relative efficiencies for the OVL measures using RSS relative to using SRS. 

\begin{table}[htbp]
\begin{center}
\caption{Asymptotic relative efficiency of OVL 
estimates using RSS relative to using SRS, $m=8$}
{\footnotesize
\begin{tabular}{|c|c|c|c|c|c|c|c|c|c|c|c|c|c|c|c|}
\hline 
 &  & \multicolumn{4}{c|}{$\rho$} &  & \multicolumn{4}{c|}{$\Delta$} &  & \multicolumn{4}{c|}{$\Lambda$} \\ 
\hline 
\multirow{2}{*}{R} & \multirow{2}{*}{\backslashbox{$r_1$}{$r_2$} } & \multirow{2}{*}{2} & \multirow{2}{*}{3} & \multirow{2}{*}{4} & \multirow{2}{*}{5} &  & \multirow{2}{*}{2} & \multirow{2}{*}{3} & \multirow{2}{*}{4} & \multirow{2}{*}{5} &  & \multirow{2}{*}{2} & \multirow{2}{*}{3} & \multirow{2}{*}{4} & \multirow{2}{*}{5} \\ 
 
 &  &  &  &  &  &  &  &  &  &  &  &  &  &  &  \\ 
\hline 
\multirow{2}{*}{0.10} & 2 & 0.9830 & 0.9795 & 0.9790 & 0.9791 &  & 0.990 & 0.9882 & 0.9879 & 0.9879 &  & 0.9998 & 0.9999 & 0.9998 & 0.9998 \\ \cline{2-16}
 
 & 3 & 0.9864&0.9839&0.9837&0.9840 &  & 0.9921&0.9908&0.9907&0.9908 &  & 0.9999&0.9999&0.9998&0.9999 \\ \cline{2-16}
 
 & 4 & 0.9882&0.9861&0.9861&0.9865 &  & 0.9932&0.9921&0.9921&0.9923 &  & 0.9999&0.9999&0.9999&0.9999 \\ \cline{2-16}
 
 & 5 & 0.9865&0.9874&0.9876&0.9880 &  & 0.9923&0.9928&0.9929&0.9932 &  & 0.9999&0.9999&0.9999&0.9999  \\ 
\hline 
 &  &  &  &  &  &  &  &  &  &  &  &  &  &  &  \\ 
\hline 
\multirow{2}{*}{0.5} & 2 &0.9021&0.9012&0.8958&0.8940  &  &  0.9814&0.9779&0.9772&0.9772 &  &  0.9828&0.9796&0.9791&0.9791 \\ \cline{2-16}
 
 & 3 & 0.9331&0.9164&0.9129&0.9125 &  & 0.9852&0.9825&0.9823&0.9825 &  & 0.9864&0.9839&0.9838&0.9840 \\ \cline{2-16}
 
 & 4 & 0.9401&0.9252&0.9228&0.9232 &  & 0.9872&0.9848&0.9849&0.9853 &  & 0.9882&0.9861&0.9862&0.9865 \\ \cline{2-16}
 
 & 5 & 0.9446&0.9308&0.9292&0.9302 &  & 0.9884&0.9863&0.9865&0.9870 &  & 0.9893&0.9875&0.9876&0.9881 \\ 
\hline 
 &  &  &  &  &  &  &  &  &  &  &  &  &  &  &  \\ 
\hline 
\multirow{2}{*}{0.8} & 2 & 0.8510&0.8036&0.7854&0.7764 &  & 0.8848&0.8523&0.8415&0.8369 &  & 0.8861&0.8541&0.8435&0.8391 \\ \cline{2-16}
 
 & 3 & 0.8647&0.8177&0.8001&0.7929 &  & 0.8991&0.8693&0.8607&0.8579 &  & 0.9004&0.8712&0.8628&0.8601 \\ \cline{2-16}
 
 & 4 & 0.8732&0.8270&0.8112&0.8045 &  & 0.9077&0.8797&0.8727&0.8711 &  & 0.9090&0.8816&0.8749&0.8733 \\ \cline{2-16}
 
 & 5 & 0.8791&0.8340&0.8188&0.8129 &  & 0.9133&0.8868&0.8809&0.8802 &  & 0.9146&0.8886&0.8829&0.8823 \\ 
\hline
\end{tabular} 
}
\end{center}
\end{table}


\begin{table}[htbp]
\begin{center}
\caption{Asymptotic relative efficiency of OVL 
estimates using RSS relative to using SRS, $m=40$}
{\footnotesize
\begin{tabular}{|c|c|c|c|c|c|c|c|c|c|c|c|c|c|c|c|}
\hline 
 &  & \multicolumn{4}{c|}{$\rho$} &  & \multicolumn{4}{c|}{$\Delta$} &  & \multicolumn{4}{c|}{$\Lambda$} \\ 
\hline 
\multirow{2}{*}{R} & \multirow{2}{*}{\backslashbox{$r_1$}{$r_2$} } & \multirow{2}{*}{2} & \multirow{2}{*}{3} & \multirow{2}{*}{4} & \multirow{2}{*}{5} &  & \multirow{2}{*}{2} & \multirow{2}{*}{3} & \multirow{2}{*}{4} & \multirow{2}{*}{5} &  & \multirow{2}{*}{2} & \multirow{2}{*}{3} & \multirow{2}{*}{4} & \multirow{2}{*}{5} \\ 
 
 &  &  &  &  &  &  &  &  &  &  &  &  &  &  &  \\ 
\hline 
\multirow{2}{*}{0.10} & 2 & 0.9931&0.9940&0.9945&0.9948 &  & 0.9961&0.9966&0.9997&0.9971 &  & 0.9999&0.9999&0.9999&0.9999 \\ \cline{2-16}
 
 & 3 &  0.9943&0.9952&0.9957&0.9960 &  & 0.9968&0.9973&0.9976&0.9978 &  & 0.9999&0.9999&0.9999&0.9999 \\ \cline{2-16}
 
 & 4 &  0.9949&0.9958&0.9963&0.9952 &  & 0.9971&0.9977&0.9979&0.9981 &  & 0.9999&0.9999&0.9999&0.9999 \\ \cline{2-16}
 
 & 5 &  0.9952&0.9962&0.9967&0.9970 &  & 0.9973&0.9979&0.9982&0.9983 &  & 0.9999&0.9999&0.9999&0.9999 \\ 
\hline 
 &  &  &  &  &  &  &  &  &  &  &  &  &  &  &  \\ 
\hline 
\multirow{2}{*}{0.5} & 2 & 0.9552&0.9603&0.9631&0.9650 &  & 0.9925&0.9935&0.9940&0.9943 &  & 0.9931&0.9940&0.9945&0.9948 \\ \cline{2-16}
 
 & 3 &  0.9621&0.9677&0.9707&0.9727 &  & 0.9937&0.9948&0.9953&0.9957 &  & 0.9943&0.9952&0.9957&0.9961 \\ \cline{2-16}
 
 & 4 &  0.9655&0.9713&0.9746&0.9767 &  & 0.9944&0.9954&0.9960&0.9964 &  & 0.9949&0.9958&0.9963&0.9967 \\ \cline{2-16}
 
 & 5 &  0.9677&0.9736&0.9770&0.9791 &  & 0.9948&0.9958&0.9964&0.9967 &  & 0.9952&0.9962&0.9967&0.9970 \\  
\hline 
 &  &  &  &  &  &  &  &  &  &  &  &  &  &  &  \\ 
\hline 
\multirow{2}{*}{0.8} & 2 & 0.8430&0.8516&0.8575&0.8616 &  & 0.9146&0.9225&0.9272&0.9304 &  & 0.9165&0.9243&0.9290&0.9321 \\ \cline{2-16}
 
 & 3 &  0.8578&0.8694&0.8772&0.8828 &  & 0.9259&0.9351&0.9406&0.9442 &  & 0.9276&0.9370&0.9421&0.9456 \\ \cline{2-16}
 
 & 4 &  0.8663&0.8798&0.8889&0.8953 &  & 0.9319&0.9419&0.9477&0.9517 &  & 0.9335&0.9433&0.9491&0.9530 \\ \cline{2-16}
 
 & 5 &  0.8718&0.8866&0.8966&0.9036 &  & 0.9357&0.9461&0.9523&0.9564 &  & 0.9373&0.9475&0.9536&0.9576 \\ 
\hline
\end{tabular} 
}
\end{center}
\end{table}

Tables 1 and 2 shows that, using $SRS$ for estimating all three overlap measure is more efficient that using $SRS$.  The efficiency increases as the set size $r_1$ and $r_2$ increases. Increasing the number of cycles's $m$ slightly decreases the efficiency. This may due the fact that this relative efficiency is based on a large sample approximation. Therefore, the larger is the sample size is the closer is the relative efficiency to the exact one.


\begin{sidewaystable} 
\caption{Bias, ratio, and length of interval (L.),  using RSS, SRS and Baye, $m=8$}
\centering
{\tiny
\begin{tabular}{|c|c|c|c|c|c|c|c|c|c|c|c|c|c|c|c|c|c|c|c|c|c|c|c|c|c|c|c|}
\hline 
 & \multicolumn{9}{|c|}{SRS}  &  \multicolumn{9}{|c|}{RSS}  & \multicolumn{9}{|c|}{BAYES} \\ 
\hline 
 &  \multicolumn{3}{|c|}{$\rho$}  &  \multicolumn{3}{|c|}{$\Delta$} & \multicolumn{3}{|c|}{$\Lambda$}  &  \multicolumn{3}{|c|}{$\rho$}  &  \multicolumn{3}{|c|}{$\Delta$} & \multicolumn{3}{|c|}{$\Lambda$} &\multicolumn{3}{|c|}{$\rho$}  &  \multicolumn{3}{|c|}{$\Delta$} & \multicolumn{3}{|c|}{$\Lambda$} \\ 
\hline   
 &$|Bias|$ & $ratio$ & $L$ & $|Bias|$ & $ratio$ & $L$ & $|Bias|$ & $ratio$ & $L$ & $|Bias|$ & $ratio$ & $L$ &$|Bias|$ & $ratio$ & $L$ & $|Bias|$ & $ratio$ & $L$ & $|Bias|$ & $ratio$ & $L$ & $|Bias|$ & $ratio$ & $L$ & $|Bias|$ & $ratio$ & $L$ \\ 
\hline
$R=.1$&&& &&& &&& &&& &&& &&& &&& &&& &&&
\\ 
$(r_1,r_2)$&&& &&& &&& &&& &&& &&& &&& &&& &&&\\
\hline
(2,2)&0.026&0.283& 0.288&0.002& 0.215&0.024 &0.001&0.024&0.146 
&0.035&0.437&0.237 &0.002&0.340&0.020 &0.001&0.040&0.120 &0.163&0.881&0.288 &0.324&0.999&0.024 &0.002&0.018&0.393
\\ 
(2,3)&0.021&0.250&0.261 & 0.013&0.193& 0.022 &0.0009&0.022&0.131 &0.027&0.392&0.208 &0.002&0.303&0.017 &0.001&0.035&0.106 &0.055&0.736&0.168 &0.110&0.999&0.014 &0.0007&0.009&0.256
\\ 
(3,3)&0.017&0.0.230&0.231 &0.001&0.173&0.019 &0.0007&0.0196&0.131 &0.019&0.337&0.175 &0.001&0.258&0.015 &0.001&0.029&0.089 &0.105&0.831&0.231 &0.208&0.999&0.019 &0.001&0.012&0.393
\\ 
(3,4)&0.014&0.214&0.214 &0.001&0.161&0.018 &0.0006&0.018&0.109 &0.017&0.317&0.163 &0.001&0.242&0.014 &0.001&0.028&0.083 &0.060&0.717&0.159 &0.098&0.999&0.103 &0.0006&0.007&0.292
\\ 
(4,4)&0.012&0.199&0.198 &0.0008&0.150&0.0170 &0.0005&0.017&0.100 &0.014&0.296&0.151 &0.0009&0.225&0.013 &0.0006&0.026&0.077 &0.077&0.789&0.198 &0.153&0.999&0.0167 &0.001&0.009&0.393
\\ 
(5,5)&0.010&0.178&0.176 &0.0006&0.133&0.015 &0.0004&0.015&0.090 &0.011&0.267&0.135 &0.0007&0.202&0.011 &0.0005&0.023&0.069 &0.061&0.752&0.176 &0.122&0.999&0.015 &0.015&0.068&0.393 \\
\hline
$R=.5$&&& &&& &&& &&& &&& &&& &&& &&& &&&
\\ 
$(r_1,r_2)$&&& &&& &&& &&& &&& &&& &&& &&& &&&\\
\hline
(2,2)&0.047&0.627&0.192 &0.020&0.295&0.212 &0.073&0.282&0.816 &0.064&0.799&0.158 &0.027&0.453&0.174 &0.099&0.436&0.671 &0.1333&0.9158&0.1923 &0.1590&0.9268&0.2121 &0.1461&0.2140&2.1931
\\ 
(2,3)&0.038&0.585&0.172 &0.159&0.266&0.190 &0.058&0.255&0.730 &0.049&0.759&0.139 &0.021&O.407&0.153 &0.076&0.391&0.589 &0.0454&0.7993&0.1122 &0.0541&08213&0.1237 &0.0497&0.1137&1.4303
\\ 
(3,3)&0.030&0.543&0.154 &0.013&0.240&0.170 &0.047&0.230&0.654 &0.035&0.699&0.117 &0.015&0.350&0.129 &0.054&0.336&0.495 &0.0857&0.8773&0.1542 &0.1022&0.8925&0.1701 &0.0931&0.1396&2.1931
\\ 
(3,4)&0.026&0.514&0.143 &0.011&0.223&0.157 &0.040&0.214&0.606 &0.030&0.675&0.109 &0.013&0.330&0.120 &0.047&0.317&0.463 &0.0405&0.7825&0.1060 &0.0483&0.8056&0.1169 &0.0044&0.0893&1.6271
\\ 
(4,4)&0.022&0.485&0.132 &0.009&0.208&0.146 &0.035&0.199&0.562 &0.026&0.646&0.101 &0.011&0.308&0.111 &0.0403&0.296&0.429 &0.0631&0.8434&0.1324 &0.0751&0.8617&0.1460 &0.0693&0.1033&2.1931
\\ 
(5,5)&0.018&0.443&0.118 &0.007&0.186&0.130 &0.027&0.178&0.500 &0.021&0.604&0.090 &0.009&0.278&0.099 &0.032&0.267&0.383 &0.0500&0.8132&0.1178 &0.0597&0.8340&0.1299 &0.0549&0.0820&2.1931 \\
\hline
$R=.75$&&& &&& &&& &&& &&& &&& &&& &&& &&&
\\ 
$(r_1,r_2)$&&& &&& &&& &&& &&& &&& &&& &&& &&&\\
\hline
(2,2)&0.043&0.851&0.086 &0.091&0.669&0.334 &0.167&0.670&0.608 &0.058&0.936&0.071 &0.126&0.829&0.275 &0.226&0.829&0.501 &0.0984&0.9661&0.0865 &0.3253&0.9545&0.3341 &0.3334&0.5572&1.6351
\\ 
(2,3)&0.034&0.823&0.077 &0.073&0.627&0.299 &0.133&0.628&0.544 &0.044&0.920&0.062 &0.096&0.793&0.242 &0.174&0.793&0.439 &0.0335&0.9092&0.0504 &0.1107&0.8817&0.1950 &0.1135&0.3306&1.0664
\\ 
(3,3)&0.027&0.793&0.069 &0.059&0.585&0.268 &0.107&0.586&0.488 &0.031&0.891&0.052 &0.067&0.738&0.203 &0.123&0.738&0.369 &0.0633&0.9487&0.0693 &0.2093&0.9319&0.2680 &0.2145&0.3962&1.6351
\\ 
(3,4)&0.023&0.769&0.064 &0.050&0.556&0.248 &0.092&0.557&0.452 &0.027&0.878&0.049 &0.059&0.715&0.189 &0.107&0.715&0.345 &0.0299&0.8999&0.0476 &0.0988&0.8701&0.1842 &0.1013&0.2649&1.2131
\\ 
(4,4)&0.020&0.745&0.060 &0.043&0.527&0.230 &0.079&0.527&0.419 &0.023&0.862&0.045 &0.0500&0.687&0.175 &0.092&0.688&0.319 &0.0467&0.9323&0.0596 &0.1543&0.9107&0.2301 &0.1581&0.3031&1.6351
\\ 
(5,5)&0.016&0.705&0.053 &0.034&0.483&0.204 &0.063&0.484&0.373 &0.019&0.836&0.041 &0.040&0.646&0.157 &0.074&0.646&0.286 &0.03697&0.9169&0.0530 &0.1222&0.8910&0.2048 &0.1252&0.2443&1.6351 \\
\hline
$R=.9$&&& &&& &&& &&& &&& &&& &&& &&& &&&
\\ 
$(r_1,r_2)$&&& &&& &&& &&& &&& &&& &&& &&& &&&\\
\hline
(2,2)&0.038&0.968&0.032 &0.344&0.942&0.405 &0.201&0.934&0.253 &0.051&0.988&0.026 &0.466&0.977&0.333 &0.272&0.974&0.208 &0.0801&0.9926&0.0322 &0.8483&0.0.9897&0.4046 &0.4022&0.8896&0.6793
\\ 
(2,3)&0.030&0.961&0.029 &0.275&0.928&0.362 &0.161&0.920&0.226 &0.040&0.984&0.023 &0.359&0.971&0.292 &0.210&0.987&0.183 &0.0273&0.9788&0.0187 &0.2892&0.9705&0.2361 &0.1369&0.7130&0.4430
\\ 
(3,3)&0.024&0.952&0.026 &0.221&0.913&0.324 &0.129&0.952&0.203 &0.028&0.978&0.019 &0.253&0.959&0.245 &0.148&0.954&0.153 &0.0515&0.9886&0.0258 &0.5463&0.9841&0.3245 &0.2587&0.7815&0.6793
\\ 
(3,4)&0.021&0.945&0.024 &0.200&0.901&0.301 &0.111&0.889&0.188 &0.024&0.975&0.018 &0.221&0.954&0.230 &0.129&0.948&0.143 &0.0243&0.9763&0.0177 &0.2580&0.9672&0.2230 &0.1222&0.6235&0.5040
\\ 
(4,4)&0.018&0.937&0.022 &0.163&0.887&0.279 &0.095&0.874&0.174 &0.021&0.971&0.017 &0.199&0.947&0.212 &0.111&0.939&0.133 &0.0379&0.9846&0.0221 &0.4027&0.9786&0.2786 &0.1907&0.6784&0.6793
\\ 
(5,5)&0.014&0.922&0.020 &0.129&0.864&0,248 &0.075&0.849&0.155 &0.017&0.964&0.015 &0.152&0.935&0.190 &0.089&0.926&0.119 &0.0300&0.9807&0.0197 &0.3189&0.9732&0.2479 &0.1510&0.5904&0.6793 \\
\hline
\end{tabular} 
}
\end{sidewaystable}

\begin{sidewaystable} 
\caption{Bias, ratio, and length of interval (L.),  using RSS, SRS and Baye, $m=40$}
\centering
{\tiny
\begin{tabular}{|c|c|c|c|c|c|c|c|c|c|c|c|c|c|c|c|c|c|c|c|c|c|c|c|c|c|c|c|}
\hline 
 & \multicolumn{9}{|c|}{SRS}  &  \multicolumn{9}{|c|}{RSS}  & \multicolumn{9}{|c|}{BAYES} \\ 
\hline 
 &  \multicolumn{3}{|c|}{$\rho$}  &  \multicolumn{3}{|c|}{$\Delta$} & \multicolumn{3}{|c|}{$\Lambda$}  &  \multicolumn{3}{|c|}{$\rho$}  &  \multicolumn{3}{|c|}{$\Delta$} & \multicolumn{3}{|c|}{$\Lambda$} &\multicolumn{3}{|c|}{$\rho$}  &  \multicolumn{3}{|c|}{$\Delta$} & \multicolumn{3}{|c|}{$\Lambda$} \\ 
\hline   
 &$|Bias|$  & $ratio$ & $L$  & $|Bias|$  & $ratio$ & $L$  & $|Bias|$  & $ratio$ & $L$  & $|Bias|$  & $ratio$ & $L$ &$|Bias|$  & $ratio$ & $L$ & $|Bias|$  & $ratio$ & $L$ & $|Bias|$  & $ratio$ & $L$  & $|Bias|$  & $ratio$ & $L$  &  $|Bias|$  & $ratio$ & $L$ \\ 
\hline
$R=.1$&&& &&& &&& &&& &&& &&& &&& &&& &&&
\\ 
$(r_1,r_2)$&&& &&& &&& &&& &&& &&& &&& &&& &&&\\
\hline
(2,2)&0.005&0.125&0.123 &0.0003&0.094&0.0104 &0.0002&0.0105&0.0628 &0.0057&0.1923&0.0956 &0.0003&0.1445&0.0080 &0.0002&0.0162&0.0486 &0.0300&0.1235&0.1235 &0.0596&0.9985&0.0104 &0.0004&0.0033&0.393
\\ 
(2,3)&0.004&0.114&0.112 &0.0002&0.085&0.009 &0.0002&0.009&0.0571 &0.0047&0.1761&0.0873 &0.0003&0.1321&0.0073 &0.0002&0.0148&0.0444 &0.0109&0.4352&0.0746 &0.0217&0.9962&0.0063 &0.0001&0.0018&0.2611
\\ 
(3,3)&0.003&0.102&0.1005 &0.0002&0.0765&0.008 &0.0001&0.0085&0.051 &0.0038&0.1580&0.0781 &0.0002&0.1183&0.0065 &0.0002&0.0132&0.0397 &0.0199&0.5459&0.1005 &0.0395&0.9979&0.0084 &0.0003&0.0022&0.3933
\\ 
(3,4)&0.003&0.0957&0.0939 &0.0017&0.0714&0.0079 &0.0001&0.0079&0.0477 &0.0033&0.1481&0.0730 &0.0002&0.1108&0.0061 &0.0001&0.0124&0.0371 &0.0097&0.4143&0.0702 &0.01929&0.9957&0.0059 &0.0002&0.0014&0.2943
\\ 
(4,4)&0.002&0.0887&0.0869 &0.0001&0.0662&0.007 &0.000&0.0073&0.0442 &0.0028&0.1373&0.0676 &0.0002&0.1027&0.0057 &0.0001&0.0114&0.0344 &0.0149&0.4908&0.0073 &0.0295&0.9971&0.0073 &0.0002&0.0016&0.3933
\\ 
(5,5)&0.002&0.0793&0.0777 &0.0001&0.0592&0.0065 &0.000&0.0066&0.0395 &0.0023&0.1230&0.0605 &0.0001&0.0920&0.0059 &0.0000&0.0102&0.0307 &0.0119&0.4496&0.0777 &0.0236&0.9965&0.0065 &0.0001&0.0013&0.3933 \\
\hline
$R=.5$&&& &&& &&& &&& &&& &&& &&& &&& &&&
\\ 
$(r_1,r_2)$&&& &&& &&& &&& &&& &&& &&& &&& &&&\\
\hline
(2,2)&0.009&0.327&0.0825 &0.0036&0.1312&0.0910 &0.013&0.1254&0.3501 &0.0104&0.4721&0.0639 &0.004&0.2007&0.0704 &0.0161&0.1921&0.2711 &0.024&0.6993&0.0825 &0.0293&0.7269&0.0910 &0.027&0.0403&2.1931
\\ 
(2,3)&0.007&0.300&0.0750 &0.003&0.1194&0.0830 &0.011&0.114&0.3184 &0.0087&0.439&0.0583 &0.004&0.1839&0.0643 &0.0134&0.1759&0.2475 &0.009&0.5086&0.0498 &0.0107&0.5386&0.0549 &0.009&0.0221&1.4559
\\ 
(3,3)&0.006&0.2709&0.0671 &0.024&0.1071&0.0741 &0.009&0.1023&0.2849 &0.0069&0.4007&0.0522 &0.003&0.1650&0.0575 &0.0107&0.1578&0.2213 &0.016&0.6229&0.0671 &0.0194&0.6527&0.0741 &0.018&0.0267&2.1931
\\ 
(3,4)&0.005&0.2542&0.0627 &0.0021&0.1000&0.0691 &0.008&0.0956&0.2660 &0.0061&0.3786&0.0488 &0.002&0.1546&0.0538 &0.0094&0.1478&0.2070 &0.008&0.4861&0.0469 &0.0094&0.5157&0.517 &0.009&0.0174&1.6414
\\ 
(4,4)&0.004&0.2365&0.0580 &0.0018&0.0927&0.0640 &0.007&0.0885&0.2463 &0.0052&0.3541&0.0452 &0.002&0.1434&0.050 &0.0081&0.1371&0.1917 &0.012&0.5670&0.0581 &0.0145&0.5974&0.0640 &0.013&0.0200&2.1031
\\ 
(5,5)&0.003&0.2125&0.0519 &0.0014&0.0823&0.0572 &0.005&0.079&0.2201 &0.0042&0.3208&0.0404 &0.002&0.1285&0.0446 &0.0064&0.1228&0.1714 &0.009&0.5239&0.0520 &0.0116&0.5541&0.0572 &0.011&0.0159&2.1931 \\
\hline
$R=.75$&&& &&& &&& &&& &&& &&& &&& &&& &&&
\\ 
$(r_1,r_2)$&&& &&& &&& &&& &&& &&& &&& &&& &&&\\
\hline
(2,2)&0.008&0.5710&0.0371 &0.0168&0.3605&0.1434 &0.031&0.3609&0.2610 &0.0094&0.7328&0.0287 &0.020&0.5136&0.1110 &0.0368&0.5140&0.2021 &0.018&0.8489&0.0371 &0.0599&0.8086&0.1434 &0.061&0.1226&1.6351
\\ 
(2,3)&0.006&0.5346&0.0338 &0.0139&0.3316&0.1304 &0.025&0.3319&0.2374 &0.0078&0.7011&0.0262 &0.017&0.4795&0.1014 &0.0307&0.4799&0.1845 &0.007&0.6962&0.0224 &0.0218&0.6386&0.0866 &0.022&0.0677&1.0855 
\\ 
(3,3)&0.005&0.4926&0.0302 &0.011&0.3001&0.1167 &0.0203&0.3003&0.2124 &0.0062&0.6604&0.0235 &0.0134&0.4391&0.0907 &0.0245&0.4395&0.1650 &0.0120&0.7942&0.0302 &0.0397&0.7455&0.1167 &0.0407&0.0815&1.6351
\\ 
(3,4)&0.004&0.4673&0.0282 &0.0097&0.2818&0.1089 &0.0177&0.2821&0.1983 &0.0055&0.6353&0.0219 &0.0117&0.4158&0.0848 &0.0215&0.4162&0.1543 &0.0059&0.6745&0.0211 &0.0194&0.6158&0.0815 &0.0199&0.0533&1.224
\\ 
(4,4)&0.004&0.4396&0.0261 &0.0083&0.2624&0.1009 &0.0152&0.2627&0.1837 &0.0047&0.6059&0.0203 &0.0101&0.389&0.0785 &0.0184&0.3901&0.1429 &0.0089&0.7489&0.0261 &0.0297&0.6952&0.1001 &0.004&0.0611&1.6351

\\ 
(5,5)&0.004&0.4396&0.0261 &0.0083&0.2624&0.1010 &0.0152&0.2627&0.1837 &0.0038&0.5630&0.0182 &0.0081&0.3540&0.0702 &0.0147&0.3544&0.1278 &0.0072&0.7106&0.0233 &0.0236&0.6538&0.0902 &0.0243&0.04879&1.635 \\
\hline
$R=.9$&&& &&& &&& &&& &&& &&& &&& &&& &&&
\\ 
$(r_1,r_2)$&&& &&& &&& &&& &&& &&& &&& &&& &&&\\
\hline
(2,2)&0.007&0.8576&0.0137 &0.0633&0.7682&0.1736 &0.037&0.7469&0.1084 &0.0084&0.9325&0.0107 &0.0759&0.8806&0.1344 &0.0444&0.8669&0.0840 &0.0300&0.6249&0.1235 &0.0596&0.9986&0.0104 &0.0004&0.0033&0.3933
\\ 
(2,3)&0.006&0.8348&0.0125 &0.0524&0.7373&0.1579 &0.0306&0.7146&0.0986 &0.0070&0.9205&0.0097 &0.0633&0.8615&0.1227 &0.03670&0.8461&0.0766 &0.0110&0.4352&0.7458 &0.0217&0.9962&0.0063 &0.0001&0.0018&0.2611
\\ 
(3,3)&0.005&0.8051&0.0112 &0.0419&0.6987&0.1413 &0.0245&0.6747&0.0882 &0.0056&0.9035&0.0087 &0.0506&0.8350&0.1098 &0.0296&0.8177&0.0685 &0.0199&0.5459&0.1005 &0.0395&0.9979&0.0084 &0.0003&0.0022&0.3933
\\ 
(3,4)&0.004&0.7850&0.0105 &0.0366&0.6738&0.1319 &0.0214&0.6492&0.0824 &0.0049&0.8919&0.0082 &0.0443&0.8175&0.1027 &0.0259&0.7989&0.0641 &0.0097&0.4143&0.0702 &0.0193&0.9956&0.0059 &0.0001&0.0014&0.2943
\\ 
(4,4)&0.003&0.7611&0.0097 &0.0314&0.6451&0.1222 &0.0183&06201&0.0763 &0.0042&0.8771&0.0075 &0.0379&0.7958&0.0951 &0.0222&0.7759&0.0594 &0.0149&0.4908&0.0869 &0.0295&0.9972&0.0073 &0.0002&0.0016&0.3932
\\ 
(5,5)&0.003&0.7237&0.0087 &0.0250&0.6023&0.1091 &0.0146&0.5769&0.0682 &0.0033&0.8528&0.0067 &0.0304&0.7616&0.0850 &0.0177&0.7399&0.0531 &0.0058&0.9111&0.0087 &0.0618&0.8811&0.1091 &0.0001&0.1403&0.6793 \\
\hline
\end{tabular} 
}
\end{sidewaystable}

\begin{figure}
\begin{center}
\label{exponentiel}
\includegraphics[scale=0.7]{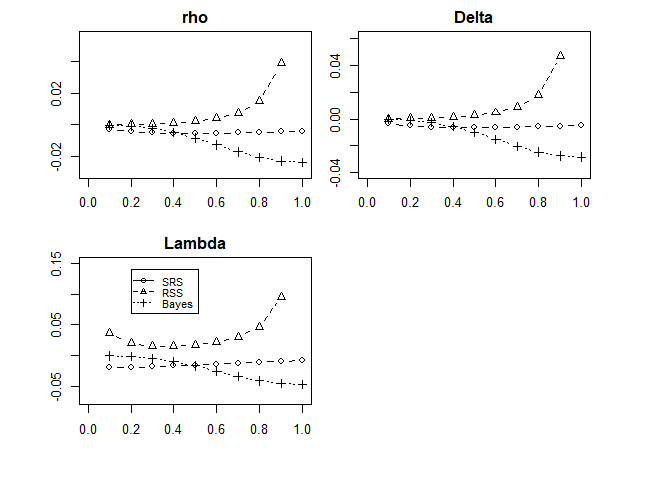}
\end{center}
\caption{The bias estimates of overlap coefficients by $R$.}
\end{figure}

\begin{figure}
\begin{center}
\label{exponentiel}
\includegraphics[scale=0.7]{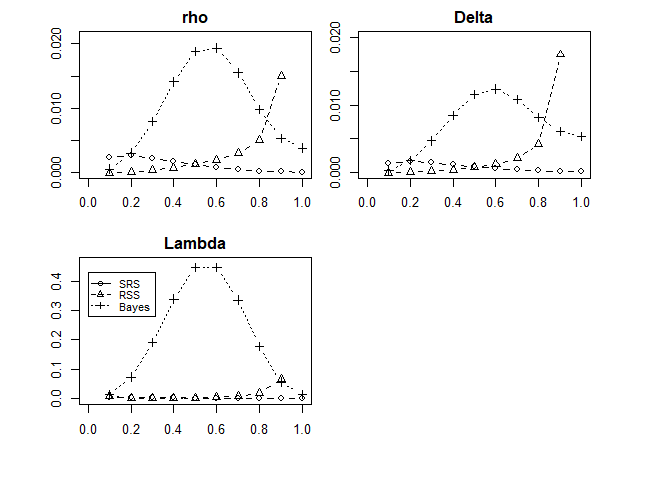}
\end{center}
\caption{The MSE estimates for overlap coefficients by $R$.}
\end{figure}

Tables 3-4 indicate that the bias of the proposed OVL estimators is negligible in most cases and $|bias|$ decreases as the sample sizes are increased for both $SRS$, $RSS$ and $Bayes$.  However, the asymptotic bias when using $SRS$ is smaller than when using $RSS$ or $Bayes$.\\
The bias estimates for $n=25$ are plotted in Figure 3. Only one plot of bias values is presented because a similar pattern is observed for other sample sizes.  For $R<0.5$ the bias estimates of the SRS, bayesian and  behave more similarly, but for the bias of RSS shows a different pattern. For $R>0.5$,   the bias estimate of the RSS is growing, that of of Bayes are decreasing and  but for that SRS tends towards 0. The estimates of MSE are plotted in Figure 4 for all three methods. For $R < 0,6$, the MSE estimates for the SRS and RSS have almost the same values  and for BB has a peak at $R = 0.6$ and declining steadily thereafter as $R$ increases.






\section{REAL DATA APPLICATION}
As applications, considers the dataset discussed by Proschan \cite{Proschan}.
The data of 30 and 12 successive failure time intervals (in hours) of the air-conditioning system of jet plane, Plane 8044 and Plane-7912, for fitting to Lomax distribution (Gupta et al. \cite{Gupta}). The inverse Lomax random variable ($X$) can be obtained by using the transformation $x=1/y$ on Lomax random variable ($Y$) (Saleem et al., \cite{Saleem}).\\
 Plane 8044: $X_1 (n = 12): 487, 18, 100, 7, 98, 5, 85,91, 43, 230, 3, 130.$ \\
Plane 7912: $X_2 (m = 30): 23, 261, 87, 7, 120, 14, 62, 47, 225, 71, 246, 21, 42,  20, 5, 12, 120, 11,$ \\
$ 3, 14, 71, 11, 14, 11, 16, 90, 1, 16, 52, 95$.\\
Fitting both data sets to  inverse Lomax distribution with parameters $\alpha_1$ (Plane 8044) and $\alpha_2$ (Plane-7912), we obtain: $\widehat{\alpha}_1= 0.0035$ and $\widehat{\alpha}_1= 0.0071$. The estimate of the ratio $\widehat{R}$ is given as $\widehat{R}=\frac{\widehat{\alpha}_1}{\widehat{\alpha}_2}=0.493$ (Table 5).\\
Since the confidence interval obtained by $RSR$ does not include the value 1 the failure time distributions for the two jets should not be considered to be identical, unlike other methods.

\begin{table}[htbp]
\begin{center}
\caption{Results based on the real data}
{\scriptsize
\begin{tabular}{|c|c|c|c|c|c|c|c|c|c|}
\hline 
 &  \multicolumn{3}{|c|}{$\widehat{\rho}=0.995$}  &  \multicolumn{3}{|c|}{$\widehat{\Delta}=0.906$}  &  \multicolumn{3}{|c|}{$\widehat{\Gamma}=0.938$}  \\ 
\hline
 & $SRS$ & $RSS$ & $Baye$ &  $SRS$ & $RSS$ & $Baye$  &  $SRS$ & $RSS$ & $Baye$  \\  
\hline  
$Bias(\widehat{OVL})$ & 0.011 &0.060  &0.012  &0.055  & 0.228  &0.046  &0.047  & 0.220  &0.040  \\ 
\hline 
$Var(\widehat{OVL})$ &0.004  &0.0003  &0.0001  &0.0015  &0.007  & 0.001 &0.022  & 0.018  &0.030  \\ 
\hline 
$95\%$ confidence & (0.991, 1.0) & (0.915 , 1.0)  &(0.990 , 1.0)  &(0.803 , 0.932)  & (0.799, 0.999) &(0.798 , 0.921)  &(0.763, 1.0)  & (0.94 , 1.0) &(0.70 , 1.0)  \\  
\hline 
\end{tabular}
} 
\end{center}
\end{table}

\section{Conclusion}
In this paper we considered three measures of overlap, namely Matusia's measure $\rho$, Weitzman's measure $\Delta$ and Kullback-Leibler $\Lambda$. We studied the estimation of overlap measures and bias and variance of their estimates. The values of the $OVL$ measures are very similar, the coefficient $\rho$ is of the best for having small values of $Bias$ and $MSE$.
The overall conclusion is that the biases of each of the $OVL$ measures are close to zero and approximations are adequate for samples of size as small as $40$.
The SRS and RSS procedures provided sensible and reasonably reliable confidence intervals. These are also the simplest methods to use in practice that do not need any computers, special software or extensive computations.

\end{document}